\documentclass[10pt, conference]{IEEEtran}
\usepackage{cite}
\ifCLASSINFOpdf
\usepackage[pdftex]{graphicx}
\else
\fi
\usepackage{url}
\usepackage{flushend}
\usepackage[utf8]{inputenc}
\usepackage{german}
\usepackage[pdftex]{graphicx}
\usepackage{wrapfig}
\usepackage{color}
\usepackage{listings}
\usepackage{amsmath}
\usepackage{mathcomp}
\usepackage{array}
\usepackage[font=footnotesize,justification=raggedright,singlelinecheck=false]{caption}

\usepackage{caption}

\begin{document}

\title{Volume Raycasting mit OpenCL}

\author{\IEEEauthorblockN{Nils~Kopal}
\IEEEauthorblockA{Universität Kassel, Angewandte Informationssicherheit,\\
Pfannkuchstra{\ss}e~1, 34121~Kassel, Germany\\
nils.kopal@uni-kassel.de}}

\maketitle

\begin{abstract}
Bei dieser Seminarausarbeitung handelt es sich um eine Ausarbeitung, die für einen 3D-Modellierungskurs (Masterstudiengang Angewandte Informatik) an der Universität Duisburg-Essen im Jahr 2011 geschrieben wurde. Ich veröffentliche diese Ausarbeitung, damit interessierte Studierende oder generell an Bildverarbeitung/Rendering Interessierte sich einen ersten Eindruck über Raycasting erwerben können. Neben der Ausarbeitung wurde ein funktionierender OpenCL Raycaster entwickelt. Ein Video ist unter \cite{Kopal} verfügbar. Der vollständige Source-Code des Raycasters ist im Google Code Archive \cite{RechKopal} archiviert. Falls dies nicht mehr der Fall ist, können Interessierte mir auch gerne eine Email schreiben, damit ich den Source-Code zur Verfügung stellen kann.

Die Seminarausarbeitung bietet eine Einführung und einen Überblick über das Thema ``Volume Raycasting mit OpenCL''. Es wird gezeigt, wie mittels moderner Grafikprozessoren Volumendatensätze in Echtzeit geladen, angezeigt und manipuliert werden können. Außerdem werden grundlegende Algorithmen und Datenstrukturen, die für dieses Thema notwendig sind, vorgestellt. Es wird gezeigt, wie ein rudimentärer Raycaster mittels OpenCL aufgebaut werden kann. Desweiteren werden verschiedene Gradientenoperatoren (CentralDifference, Sobel3D und Zucker-Hummel) vorgestellt, implementiert und evaluiert. Abschließend werden noch Beschleunigungsmöglichkeiten für das Raycasting vorgestellt.
\end{abstract}

\section{Einleitung}
Volume Raycasting findet heutzutage besonders in der medizinischen Informatik aber auch im Bereich des Entertainments, hier in Computerspielen, seinen Einsatz. Beim Raycasting werden mittels Strahlenverfolgung Volumendatensätze abgetastet und aus diesen Abtastdaten ein Bild errechnet. Diese Volumendaten enthalten z.B. mittels 3D-Scanner erfasste Objekte. Im medizinischen Bereich werden mit Hilfe von Raycasting mittels Computertomographie, Magnetresonanztomographie oder ähnlichen erzeugte Volumendaten, visualisiert. In Computerspielen werden Raycastingverfahren genutzt, um realistischere Rauch-, Nebel- und Feuereffekte zu erzeugen. In dieser Ausarbeitung wird näher auf die medizinische Anwendung, also das Visualisieren von medizinischen Daten eingegangen. So werden sowohl der berühmte ``Stanford Bunny'' als auch drei medizinische Volumendatensätze, die jeweils mittels Computertomographie/Magnetresonanztomographie  erstellt wurden, durch einen auf der GPU (\textit{Graphics Processing Unit }) ausgeführten Raycaster visualisiert.

Medizinische Geräte zur Untersuchung menschlichen Gewebes, wie z.B. Computertomographen oder Magnetresonanztomographen, liefern uns eine schichtweise Darstellung des menschlichen Körpers. So erzeugen solche Geräte eine Vielzahl von Schnittbildern, die aneinandergereiht, einen vollständigen menschlichen Körper modellieren. Weist man den einzelnen Messpunkten innerhalb dieser ``Bilder'' Graustufen oder Farbwerte hinzu, so erhält man eine relativ genaue Visualisierung, jedoch nur in Schichten. Diese Schichten lassen sich vorwärts und rückwärts durchlaufen, bieten aber kein Gesamtbild des Körpers oder einzelner Teile. Nutzt man diese Daten jedoch als Gesamtes, so ermöglicht das Raycasting eine dreidimensionale originalgetreue Abbildung des gesamten Körpers oder, wenn notwendig, sogar nur einzelner Bereiche wie Organe, Knochen etc. Medizinischem Personal ist es so möglich, eine 360 Grad Ansicht der abgescannten Person einzusehen, ``unter die Haut'' zu dringen um so medizinische Maßnahmen ohne direkten Eingriff am Menschen zu planen.

Dafür notwendige Raycasting-Verfahren sind relativ Prozessorlastig, jedoch einfach zu parallelisieren. Das ermöglicht es Informatikern die Algorithmen auf moderne Grafikkarten zu portieren um so 3D-Anwendungen zu Erstellen, die den menschlichen Körper in Echtzeit darstellen.

Grafikkarten waren in der Vergangenheit zunächst als Entlastung für die CPU (\textit{Central Processing Unit}) des Computers gedacht. Sie boten fest eingebaute Funktionen um zweidimensionale, gerasterte Grafiken zu manipulieren. Im Zuge der Entwicklung von 3D-Grafiken verfügten Grafikkarten mehr und mehr über Funktionen um auch diese zu Erzeugen. So dienen Grafikkarten dazu, den Prozessor bei der Projektion und Texturierung sowie Beleuchtung von 3D-Szenen zu unterstützen oder ihm diese Arbeit sogar gänzlich abzunehmen. Im Laufe dieser Entwicklung verfügten Grafikkarten über immer mehr Operationen welche in Standards wie OpenGL oder DirectX zusammengefasst wurden. Grafikkarten trugen zu einer deutlichen Steigerung der Rendergeschwindigkeit bei und ermöglichten somit immer umfangreichere und komplexere Szenen.

Bald schon ermöglichten sogenannte \textit{Shader} die Manipulation der Grafikpipeline in bestimmten Abschnitten. Bei einem Shader handelt es sich um ein kleines Programm das direkt in der Grafikkarte ausgeführt wird. Dies ermöglicht dem Entwickler den Renderingprozess für seine Anforderungen individuell anzupassen.

Der neueste Trend bei den Grafikkarten ist heute die vollständige Programmierung dieser. So ist es nicht mehr nur möglich, einzelne Schritte der Renderpipeline anzupassen, sondern mittels Grafikprogrammierung die komplette Grafikkarte zu steuern. Hier sind vor allem NVidias CUDA und der offene Standard OpenCL zu nennen, die diese direkte Programmierung ermöglichen. Mittlerweile ermöglicht dies auch Algorithmen auf Grafikkarten zu portieren, die eigentlich mit Grafikprogrammierung überhaupt nichts mehr zu tun haben (wie z.B. das Brechen von Verschlüsselungen mit Hilfe der Grafikkarte). Nun können aber auch Raytracing- und Raycasting-Verfahren vollständig in einer dieser Sprachen umgesetzt und vollständig von der Grafikkarte berechnet werden. Da Grafikkarten für Berechnungen dieser Art hoch optimiert sind erreichen diese Algorithmen auf der Grafikkarte bisher nie dagewesene Rechengeschwindigkeiten und ermöglichen sogar deren Darstellung in Echtzeit.

Der Rest der Ausarbeitung ist wie folgt aufgebaut: Im Grundlagen Kapitel werden zunächst Volumendatensätze vorgestellt und es wird gezeigt, wie diese generiert und verarbeitet werden können. Die Grundlagen und die Unterschiede zwischen Raycasting und Raytracing, welche für die Anzeige der Volumendatensätze genutzt werden, werden in diesem Kapitel ebenfalls vorgestellt. Abschließend wird die Programmierplattform OpenCL (\textit{Open Computing Language}), mit der man z.B. auf NVIDIA Grafikkarten programmieren kann, kurz vorgestellt. Danach wird ein rudimentärer Raycaster vorgestellt, der im Zuge dieser Ausarbeitung entwickelt worden. Dieser wird daraufhin kurz evaluiert. Hier werden die verschiedenen implementierten Gradientenoperatoren bezüglich ihrer Leistung (in Bildern pro Sekunde) evaluiert. Dann werden verschiedene Beschleunigungsmöglichkeiten für das Raycasting vorgestellt. Den Abschluss dieser Ausarbeitung bildet eine Zusammenfassung sowie ein kurzer Ausblick in mögliche Folgearbeiten.

\section{Grundlagen}
Dieses Kapitel bietet eine kleine Einführung in die Grundlagen des Raycastings. So werden zunächst Volumendatensätze erläutert. Im zweiten Teil wird kurz auf Raytracing und Raycasting und deren Unterschiede eingegangen. Abschließend wird auf den von der Khronos Group entwickelten Standard OpenCL eingegangen, welcher benutzt werden kann, um Programme direkt für die Grafikkarte zu entwickeln.

\begin{figure}
\begin{center}
  \includegraphics[width=0.75\columnwidth]{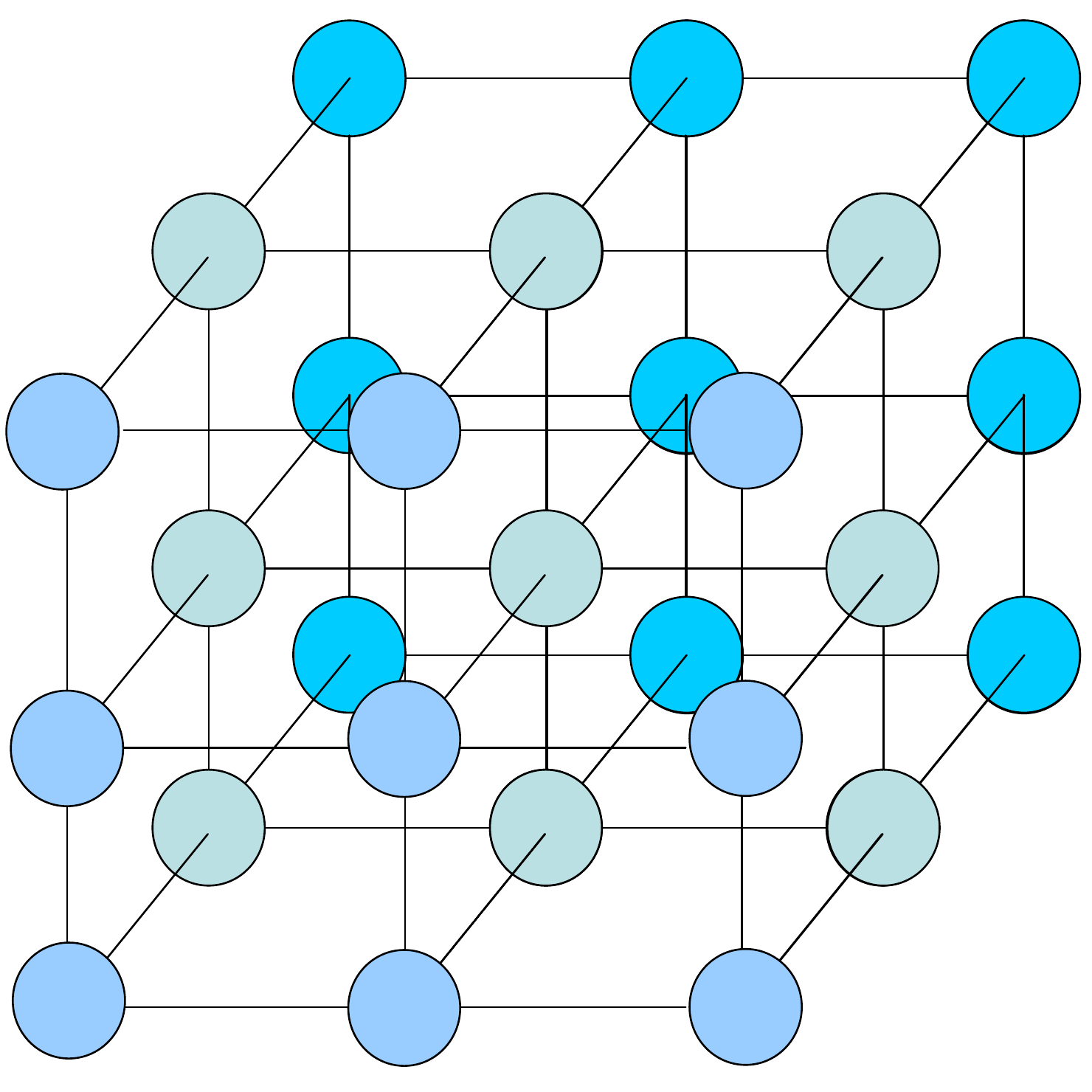}
  \caption[Schematische Darstellung eines Volumendatensatzes]{Schematische Darstellung eines Volumendatensatzes}
  \label{label:Volumendatensatz}
\end{center}
\end{figure}

\subsection{Volumendatensätze}
Ein Volumendatensatz ist eine dreidimensionale Funktion, die jedem $(x,y,z)$-Tupel einen Funktionswert zuordnet:

\begin{equation}
 V: \mathrm{N}^3 \rightarrow \mathrm{R}
\end{equation}

Man stelle sich einen Quader vor, der aus senkrecht und waagerecht angeordneten Gitternetzpunkten besteht. Jeder Gitterpunkt beinhaltet genau einen Funktionswert. Diese Punkte werden auch als Voxel bezeichnet, wobei sich der Name aus Volumen und Pixel zusammensetzt. Da Volumendatensätze nur aus diesen Punkten bestehen, müssen Funktionswerte, die zwischen mehreren Punkten liegen, interpoliert werden. Dazu wird lineare oder trilinieare Interpolation zwischen zwei oder acht Gitterpunkten verwendet. Gerade im medizinischen Umfeld werden Volumendatensätze zur Visualisierung des menschlichen Körpers genutzt. So wird zunächst eine Aufnahme eines Körpers mittels eines Computertomographen, eines Magnetresonanztomographen oder ähnlichen Geräten erstellt. Diese Geräte speichern ihre gewonnen Daten innerhalb eines Volumendatensatzes ab (im einfachsten Fall als ein dreidimensionales Array von Abtastwerten). Diese Datensätze können dann später am Computer geladen und in einer dreidimensionale Abbildung dargestellt werden. Von derlei Geräten erzeugte Volumendatensätze sind häufig sehr groß. So ist ein würfelförmiger Datensatz mit einer Kantenlänge von 512 Punkten, wobei jedes Datum in einem Short (2 byte) gespeichert würde, bereits 256 Megabyte groß ($512^3 * 2 $ byte $ = 256 $ Megabyte).

\subsection{Raytracing und Raycasting}
Raytracing ist ein Verfahren der Computergrafik, mit dem sehr realistische Computergrafiken von dreidimensionalen Szenen gerendert werden können. Beim Raytracing werden, ausgegangen vom Betrachter einer Szene, für jeden Bildpunkt Strahlen, die sogenannten Rays, losgeschickt. Trifft ein Strahl auf ein Objekt, dass sich in der Szene befindet, so werden Oberflächeneigenschaften, Beleuchtung der Szene und gegebenenfalls Reflektionen und Refraktionen für den Schnittpunkt berechnet. Aus diesen Informationen lässt sich dem Bildpunkt ein Farbwert zuweisen, der dann dem Betrachter angezeigt wird. Treffen Strahlen auf reflektierende oder refraktierende Oberflächen, werden Schnittwinkel bestimmt und Folgestrahlen ausgesandt, um festzustellen, ob Informationen weiterer Objekte mit in den aktuellen Bildpunkt eingerechnet werden müssen. Um dies zu vereinfachen, sind Raytracer häufig mittels rekursiver Algorithmen implementiert. Da ein einfaches Bild mit einer Auflösung von 640 * 480 Bildpunkten bereits 307.200 Strahlen vom Betrachter aussendet, sind Raytracer in ihrer Berechnungsgeschwindigkeit häufig sehr langsam. Die Hauptrechenlast bei Raytracern wird durch die Schnittpunktberechnungen zwischen Strahlen und in der Szene angeordneten Objekten erzeugt. So muss im allgemeinen Fall, ohne jede Optimierung, für alle Strahlen ein möglicher Schnittpunkt mit jedem Objekt der Szene berechnet werden. Ist ein Schnittpunkt gefunden, wird die Beleuchtung des Punktes, z.B. mittels Phong Shading berechnet. Hierfür berechnet man auch den Normalenvektor im Schnittpunkt der getroffenen Oberfläche und nutzt diesen um besagtes Phong Shading durchzuführen.

Im Gegensatz zum Raytracing werden beim Raycasting Strahlen nicht gegen Objekte, wie z.B. Kugeln, Würfel oder Ebenen geleitet. Beim Raycasting werden Strahlen auf einen der oben beschriebenen Volumendatensätze geleitet. Trifft ein solcher Strahl auf einen Voxel, so wird aus dem Schnittpunkt, dem Auftreffwinkel und dem Wert an entsprechende Volumenposition ein Farbwert bestimmt, der dem Betrachter dann angezeigt wird. Im Gegensatz zum Raycasting wird eine einfache Beleuchtung, bestehend aus nur einer Lichtquelle, genutzt. Außerdem werden keine Reflektions- und auch keine Refraktionsstrahlen ausgesandt. Häufig werden auch Strahlen, nach dem Aufprall auf ein Voxel, weiter innerhalb des Volumendatensatz verfolgt. Die ``tiefer'' liegenden Voxel werden so auch mittels eines Emissions-Absorptions-Modell in den Farbwert mit eingerechnet. Dies ermöglicht dem Betrachter, z.B. in medizinischen Anwendungen, auch innerhalb eines Körpers liegende Regionen des Volumendatensatzes, z.B. Adern oder Knochen, die unter der Haut liegen, zu betrachten. Hierfür werden physikalische Grundlagen genutzt. So wird die Energie des Lichtes auf seinem Weg durch den Volumendatensatz ``immer mehr aufgebraucht'' und das Licht schließlich von Knochen vollständig absorbiert. So kann man durch dünne Hautschichten hindurch sehen, Knochen stoppen die Strahlen jedoch.

Da ein Voxel genau genommen keine räumliche Ausdehnung besitzt und auch keine Oberfläche, ist die Bestimmung des Auftreffwinkels eines Strahles nicht direkt möglich. Abhilfe schaffen hier Bildableitungsfunktionen die den Gradienten des Voxels bestimmen. Im nächsten Kapitel wird auf diese Gradientenfunktionen näher eingegangen.

Der große Vorteil sowohl von Raytracing als auch Raycasting sind ihre einfache parallelisierbarkeit. Da für jeden Bildpunkt ein eigener Strahl verfolgt wird, der absolut unabhängig von denen andere Bildpunkte ist, kann man jeden Strahl separat berechnen. So lässt sich ein Raytracer/Raycaster schon auf einem Multiprozessor-System deutlich beschleunigen, indem man die Berechnungen auf mehrere der Prozessoren des Systems verteilt. So halbiert sich die Berechnungsgeschwindigkeit bereits beim hin zuschalten eines zweiten Prozessors. Moderne Grafikkarten verfügen Heutzutage zwischen 16 (Geforce 8400 GS) und 128 (Geforce 8800 Ultra) und sogar bis zu 480 (Geforce GTX 480) sogenannter Stream Prozessoren. So können parallele Algorithmen massiv parallel ausgeführt werden. Im Fall von Raytracing/Raycasting werden nun Bildwiederholraten durch Portierung der Algorithmen (mittels CUDA oder OpenCL) auf die Grafikkarte von mehren Bildern pro Sekunde möglich.

\subsection{OpenCL}
OpenCL, auch Open Computing Language, ist ein von der Khronos Group entwickelter Standard. Mit diesem und der Sprache OpenCL C können Programme sowohl für CPUs als auch für GPUs (Grafikprozessoren) entwickelt werden. OpenCL Programme (auch Kernels genannt) können während der Ausführung auf mehrere OpenCL-fähige Geräte verteilt werden. Ein OpenCL-System besteht immer aus zwei oder mehreren Komponenten. Zum einem dem sogenannten Host-Programm und aus mehreren OpenCL fähigen Devices. Ein Device kann z.B. eine CPU oder eine GPU darstellen. Die Devices wiederum bestehen aus einer oder mehreren Computing Units. Bei CPUs sind dies die Kerne und bei Grafikkarten die Streamprozessoren. Zur Laufzeit werden die Kernels (OpenCL Programme) auf diese Streamprozessoren durch den Host verteilt und können somit parallel abgearbeitet werden. OpenCL Kernel werden erst zur Laufzeit in ausführbaren Code übersetzt. Somit kann ein in OpenCL C geschriebenes Programm auf unterschiedlichen Zielplattformen ausgeführt werden und es muss zur Entwicklungszeit nicht feststehen, was die Zielplattform ist. So übersetzt der NVidia Compiler das OpenCL C Programm zunächst in CUDA-Code um es dann mittels CUDA auszuführen. OpenCL wurde für die Entwicklung von massiv parallelen Anwendungen konzipiert. So kann z.B. ein Computerbild der Auflösung 1024 x 1024 komplett parallel verarbeitet werden: 

\begin{small}
\begin{lstlisting}[caption=Klassisches C Programm, label=label:klassischC]{Klassisches C Programm}
 void trad_mul(int n,
               const float *a,
               const float *b,
               float *c)
 {
    int i;
    for(i = 0; i < n; i++)
 	{
 		c[i] = a[i] * b[i];
 	}// klassische schleife
 }
\end{lstlisting}

\begin{lstlisting}[caption=OpenCL C Programm, label=label:openCL]{OpenCL C Programm}
 kernel void trad_mul(global const float *a,
                      global const float *b,
                      global float *c)
 {
 	int id = get_global_id(0);
 	c[id] = a[id] * b[id];
 } // fuehre n work items parallel aus
\end{lstlisting}
\end{small}

Wie in den Codebeispielen \ref{label:klassischC} und \ref{label:openCL} (entnommen aus \cite{openclintro}) zu erkennen, kann man klassische Schleifen relativ einfach in einen OpenCL Kernel umschreiben um diesen dann parallel auszuführen. Im Codebeispiel \ref{label:klassischC} werden zwei Computerbilder multipliziert und das Ergebnis in einem dritten Bild gespeichert. Dies geschieht mit Hilfe einer einfachen for-Schleife die alle Operationen sequenziell abarbeitet. Im Codebeispiel \ref{label:openCL} wird das selbe Ergebnis mit Hilfe eines OpenCL Kernels erzielt. Mittels des Aufrufs \textit{get\_global\_id(0);} erhält das aktuell ausgeführte Work Unit seine ID und kann daraus die Position in den Bildarrays errechnen, die es zu bearbeiten hat. Im Gegensatz zum klassischen C-Programm kann der OpenCL Kernel von einer Vielzahl von Work Units parallel abgearbeitet werden.\\
\\
OpenCL verfügt über eine Vielzahl von Datentypen und Funktionen, die Grafikprogrammierung unterstützen und vereinfachen. So existieren die Datentypen \textit{float2, float3, float4, float8} und \textit{float16} im OpenCL Standard. Diese entsprechen n-dimensionalen Vektoren deren Komponenten \textit{float} Werte beinhalten. So kann eine Variable x des Typs \textit{float3} z.B. über seine Komponenten \textit{x.S0}, \textit{x.S1} und \textit{x.S2} angesprochen werden. Arithmetische Operationen auf Vektor-Datentypen werden von den GPUs unterstützt und entsprechend schnell berechnet. So ist eine Addition zweier \textit{float4}- Werte auf einer GPU ebenso schnell oder gar schneller wie die Addition zweier reiner \textit{float}-Werte mittels einer CPU. OpenCL liefert, wie bereits erwähnt, auch mehrere nützliche Funktionen für die Grafikprogrammierung. So kann mittels \textit{fast\_normalize(x)} ein \textit{float-n}-Wert schnell normalisiert werden. Ebenso werden Funktionen zur Berechnung des Skalarprodukts (\textit{dot(v1,v2)}) oder des Kreuzprodukts (\textit{cross(v1,v2)}) zweier Vektoren mitgeliefert, welche durch die GPU ebenfalls besonders schnell ausgeführt werden können.

\section{Entwurf eines Raycasters mit OpenCL}
In diesem Kapitel wird ein einfacher Raycaster mit OpenCL und .NET C\# (Hostprogramm) entworfen. Zu diesem Raycaster gehört unter anderem das Laden und Anzeigen bereitgestellter Volumendatensätze, die durch einen Computertomographen erstellt wurden. Diese Volumendatensätze sollen im Raum positioniert und mittels einer virtuellen, positionierbaren Kamera dargestellt werden. Der Raycaster ermöglicht das Manipulieren dieser Datensätze. Darunter fällt das Vergrößern und Verkleinern des Bildausschnitts. Außerdem wird es möglich sein, den Datensatz um eine Achse zu Drehen, um ihn so von mehreren Seiten betrachten zu können. Die Lichtquelle, die den Volumendatensatz erhellt, wird im Raum bewegt werden können. Außerdem kann man die sogenannte Bounding-Box um den Datensatz einstellen um diesen sozusagen zu ``zerschneiden''. Damit das Volumen auch auf schwächeren Rechnern dargestellt werden kann, wird die Render-Qualität (die Auflösung und die sogenannte ``Gradientenfuntkion'') einstellbar sein. Der Raycaster wird nur die Oberflächen der sich innerhalb des Volumen befindlichen Objektes abtasten, jedoch nicht tiefer in das ``Material'' eindringen. Die Implementierung des hier vorgestellten Raycasters kann von \cite{RechKopal} heruntergeladen werden.

\subsection{Hostprogramm}
Als Hostprogramm für den Raycaster dient ein .NET C\# Programm das die Volumendatensätze einließt und in den Grafikkartenspeicher kopiert. Die zur Verfügung stehenden Volumendatensätze bestehen aus n Binärdateien, die 512 X 512 große Rohdaten von Messwerten enthalten, die ein Computertomograph erzeugt hat. Diese werden zunächst in ein dreidimensionales Array und dann mit Hilfe der OpenCL API in die Grafikkarte kopiert. Sobald das Volumen vollständig geladen ist, werden mehrere Instanzen eines OpenCL Kernels erstellt. Diese arbeiten jeder für sich genau einen Strahl in das Volumen ab und speichern ihre Ausgabe in einem großen Bildbuffer der dann final von dem C\#-Programm wieder angezeigt wird.

\subsection{Laden eines Volumendatensatzes}
Die in diesem Raycaster verwendeten Volumendatensätze stammen aus dem \cite{TS3DSR} ``Stanford 3D Scanning Repository''. Bei den Datensätzen handelt es sich zum einem um einen CT-Scan des berühmten ``Stanford Bunny''. Dieser Porzellanhase dient in der 3D Grafikszene einer Vielzahl von Wissenschaftlern als Datenvorlage. In diesem Raycaster stellt er den ersten von drei Volumendatensätzen dar. Der ``Hase'' besteht aus 360 Binärdateien, die jeweils ein 512 X 512 großes ``Bild'', eine Scheibe, des Hasen darstellen. Im Raycaster wird ein kubischer Volumendatensatz der Kantenlänge 512 erstellt in dem die Scheiben an die Koordinaten $Z=0$ bis $Z=360$ geladen werden. Jeder Voxel besteht aus einem 16bit breiten short-Value, der jedoch nur 12bit Daten enthält. Nur 12bit aus dem Grund, dass CT-Scanner Daten, basierend auf der Hounsfield-Skala, liefern. Für diese Skala reichen 12bit vollkommen aus. Neben dem Bunny ermöglicht der Raycaster das Laden und Anzeigen eines CT-Scans eines menschlichen Kopfes sowie eines CT-Scans eines weiteren Kopfes, dem die Schädeldecke medizinisch entfernt wurde, damit das Gehirn frei liegt. Der letzte Volumendatensatz ist ein menschlicher Oberkörper, bei dem sowohl die Knochen als auch die Organe ``freigelegt'' werden können.

\begin{figure}
\begin{center}
  \includegraphics[width=0.75\columnwidth]{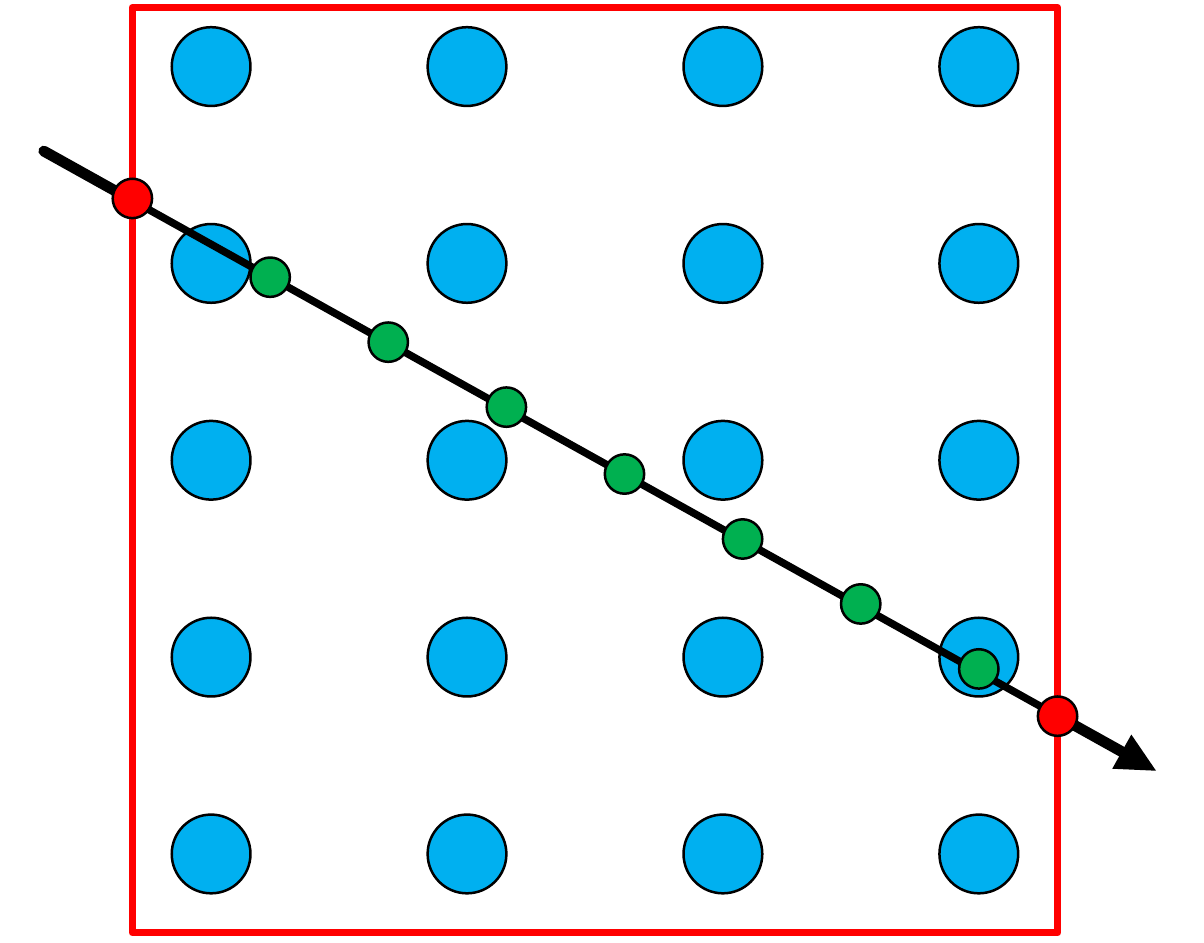}
  \caption[Volumenschnittpunkt/Abtastung]{Volumenschnittpunkte/Abtastung}
  \label{label:VolumenschnittpunktAbtastung}
\end{center}
\end{figure}

\subsection{Schnittpunktbestimmung/Abtastung im Volumen}
Zunächst wird der Schnittpunkt des zu berechnenden Strahls und einer Bounding Box berechnet. Die Bounding Box umschließt den Volumendatensatz. Gibt es keinen Schnittpunkt, so wird als Farbe Schwarz ausgeben. Existieren zwei Schnittpunkte, so wird zwischen diesen beiden Punkten entlang des Strahls ``abgetastet''. Der Abstand zwischen den Abtastpunkten spiegelt sich später sowohl in der Qualität das erzeugten Bildes als auch in der Berechnungsgeschwindkeit wieder. Da Volumendatensätze aus einzelnen Punkten bestehen und die Wahrscheinlichkeit, dass ein Strahl exakt auf einen Punkt trifft, sehr gering ist, müssen die Werte zwischen mehreren Punkten interpoliert werden. Als einfachstes Verfahren (aber auch qualitativ nicht sehr hochwertig) lassen sich die Koordinaten auf ganzzahlige Werte Runden. Alternativ kann man auch zwischen zwei Punkten linear oder zwischen 8 Punkten trilinear interpolieren. Dies liefert qualitativ höherwertige Bilder.

In Abbildung \ref{label:VolumenschnittpunktAbtastung} ist eine Abtastung eines Volumendatensatzes schematisch dargestellt. Der schwarze Pfeil stellt den Strahl dar, der durch das Volumen gecastet wird. Zwischen dem Eintrittspunkt und dem Austrittspunkt des Strahls und der Bounding-Box (hier in rot dargestellt) werden in regelmäßigen Abständen Punkte abgetastet (hier in grün dargestellt). Hier wird auch deutlich, warum zwischen den eigentlichen Voxeln (blaue Punkte) weitere Funktionswerte interpoliert werden müssen, da der Strahl genau genommen keinen der Voxel genau trifft.

\begin{figure}
\begin{center}
  \includegraphics[width=0.75\columnwidth]{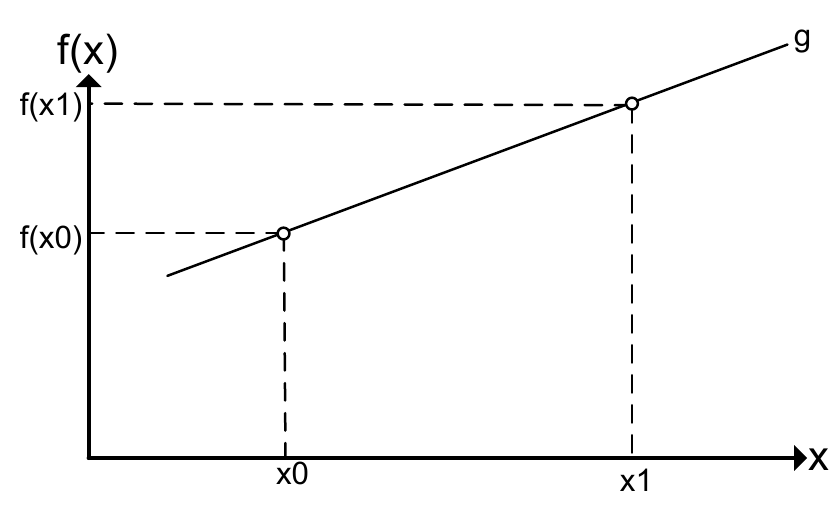}
  \caption[Lineare Interpolation]{Lineare Interpolation}
  \label{label:LineareInterpolation}
  \includegraphics[width=0.75\columnwidth]{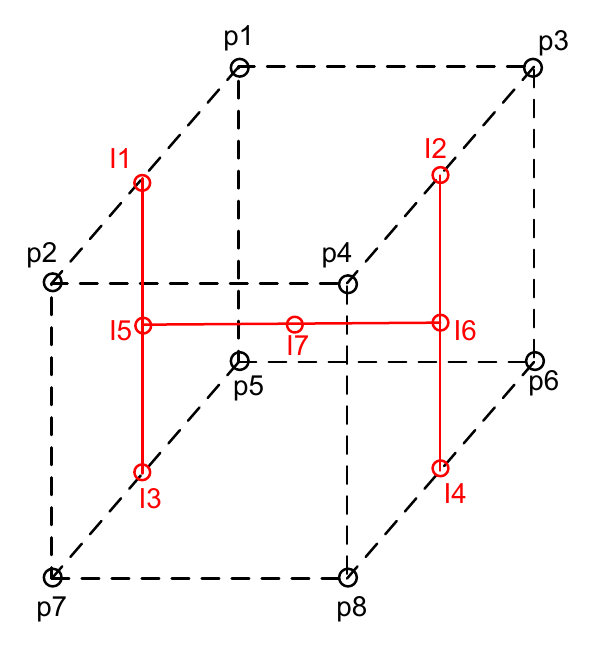}
  \caption[Trilineare Interpolation]{Trilineare Interpolation}
  \label{label:TrilineareInterpolation}
\end{center}
\end{figure}

\subsection{Lineare und trilineare Interpolation}
Um ein Datum aus dem Volumensatz zu berechnen, das zwischen einzelnen Voxeln liegt, wird lineare und trilineare Interpolation genutzt. Bei der linearen Interpolation wird eine einfache Gerade $g$ zwischen zwei Funktionswerte $f(x0)$ und $f(x1)$ gelegt um Zwischenwerte zu berechnen. So kann mittels Einsetzen eines Wertes in die Geradengleichung ein Zwischenwert interpoliert werden. Mittels der Formel

\begin{equation}
	g(x) = f(x0) + \frac{f(x1)-f(x0)}{x1 - x0} ( x - x0)
\end{equation}

kann die Geradengleichung bestimmt werden. In Abbildung \ref{label:LineareInterpolation} ist das Verfahren als Skizze verdeutlicht. Im dreidimensionalen wird die trilineare Interpolation bei den Voxeln angewandt. Hierfür werden die diskreten Werte von acht Voxeln in Zweierpaaren linear interpoliert. In Abbildung \ref{label:TrilineareInterpolation} werden zunächst die Paare $(p1,p2), (p3,p4), (p5,p6), (p7,p8)$ mittels linearer Interpolation miteinander kombiniert. Anschließend werden deren Ergebnispaare $(I1, I3), (I2, I4)$ nochmals linear interpoliert. Abschließend erhält man den gesuchten Funktionswert $I7$ nach dritter linearer Interpolation des Paares $(I5,I6)$.

\subsection{Gradientenfunktion}
Hat der Strahl im Volumen einen Punkt erreicht, dessen Wert ungleich 0 (= leerer Raum) ist, so kann diesem Punkt ein Farbwert zugewiesen werden (Shading). Da für die Beleuchtung im dreidimensionalen Raum (mittels Phong-Shading) allerdings ein Normalenvektor notwendig ist, muss dieser zunächst berechnet werden. Um aus einem Volumendatensatz Normalenvektoren zu berechnen bieten sich, aus der klassischen Bildverarbeitung bekannte, Verfahren an. Anstelle des Normalenvektors nutzt man die Ableitung der Bildfunktion bzw in unserem Fall der Volumen-Funktion, was dem Gradienten entspricht. Hierfür nutzt man einfache oder auch komplexere Gradientenfunktionen die sich der Bildableitung relativ genau annähern. Als sehr einfache Gradientenfunktion ist der \textit{Central-Difference-Operator} zu gebrauchen. Dieser bestimmt die lokalen Ableitungen in X, in Y und in Z-Richtung indem er folgende Rechnungen vornimmt:

\begin{equation}
  \begin{split}
	\delta_X = V(P.X + 1,P.Y,P.Z) - V(P.X - 1,P.Y,P.Z) \\
	\delta_Y = V(P.X,P.Y + 1,P.Z) - V(P.X,P.Y - 1,P.Z) \\
	\delta_Z = V(P.X,P.Y,P.Z + 1) - V(P.X,P.Y,P.Z - 1) \\
	G = normalize( (\delta_X, \delta_Y, \delta_Z) )
  \end{split}
\end{equation}

\begin{figure}[htbp]
  \begin{center}
  \includegraphics[width=0.5\columnwidth]{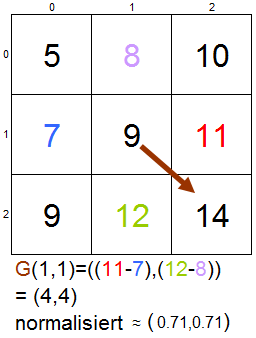}
  \caption[Beispielrechnung Central-Difference-Operator]{Beispielrechnung Central-Difference-Operator}
  \label{label:gradient}
  \end{center}
\end{figure}

\begin{figure}[htbp]
  \begin{center}
  \includegraphics[width=0.5\columnwidth]{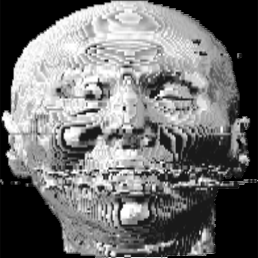}
  \caption[Kopf Central-Difference-Operator]{Kopf Central-Difference-Operator}
  \label{label:kadaver_central_difference}
  \includegraphics[width=0.5\columnwidth]{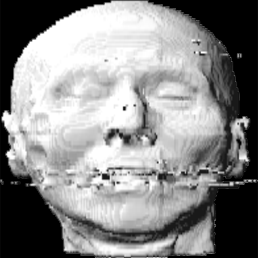}
  \caption[Kopf Zucker-Hummel-Operator]{Kopf Zucker-Hummel-Operator}
  \label{label:kadaver_zucker_hummel}
  \end{center}
\end{figure}

Zunächst berechnet man die jeweiligen Funktionswerte des Volumens $V(x,y,z)$ in allen drei Achsenrichtungen vor dem Punkt P minus den entsprechenden Funktionswerten in gleicher Achsenrichtung nach dem Punkt P. Der neu berechnete Vektor $(\delta_X, \delta_Y, \delta_Z)$ wird abschließend noch normalisiert (durch seinen Betrag geteilt) und ergibt so den Gradienten. Der Gradient zeigt somit immer in die Richtung, in der die größten Funktionswerte stehen. Als Beispiel im zweidimensionalen ist die Abbildung \ref{label:gradient} beigefügt, die eine Gradientenbestimmung in X und in Y-Richtung zeigt.\\
\\
Mit Hilfe des \textit{Central-Difference-Operator} lassen sich relativ einfach und vor allem relativ schnell Gradienten im Volumen bestimmen. Um qualitativ bessere Ergebnisse zu erhalten bietet sich bessere Operatoren wie z.B. der \textit{Zucker-Hummel-Operator} oder der \textit{Sobel 3D Operator} an. Sowohl der Zucker-Hummel als auch der Sobel3D-Operator nutzen für die Gradientenberechnung, im Gegensatz zum Central-Difference-Operator nicht nur 6 Voxel sondern insgesamt 54 Voxel. Dadurch fallen die Gradienten deutlich weicher und klarer aus als dies beim Central-Difference-Operator der Fall ist. Der qualitative Unterschied zwischen dem \textit{Central-Difference-Operator} und dem \textit{Zucker-Hummel-Operator} wird in den Abbildungen \ref{label:kadaver_central_difference} und \ref{label:kadaver_zucker_hummel} deutlich.\\
\\

Als komplexeres Beispiel für einen Gradientenoperator ist in Formel~\ref{Sobel3D} der Sobel3D-Operator mittels folgender Operatormasken dargestellt (entnommen von \cite{S01Project}).
\begin{figure*}
\begin{equation}
 \begin{split}
	x-1:
	\begin{bmatrix}
	\hphantom{\hspace{20pt}} & \hphantom{\hspace{20pt}} & \hphantom{\hspace{20pt}}\\[-15pt]
	-1 & -3 & -1\\
	-3 & -6 & -3\\
	-1 & -3 & -1
	\end{bmatrix}
	x:
	\begin{bmatrix}
	\hphantom{\hspace{20pt}} & \hphantom{\hspace{20pt}} & \hphantom{\hspace{20pt}}\\[-15pt]
	\ 0 & \ 0 & \ 0\\
	\ 0 & \ 0 & \ 0\\
	\ 0 & \ 0 & \ 0
	\end{bmatrix}
	x+1:
	\begin{bmatrix}
	\hphantom{\hspace{20pt}} & \hphantom{\hspace{20pt}} & \hphantom{\hspace{20pt}}\\[-15pt]
	\ 1 & \ 3 & \ 1\\
	\ 3 & \ 6 & \ 3\\
	\ 1 & \ 3 & \ 1
	\end{bmatrix}
	\\
	y-1:
	\begin{bmatrix}
	\hphantom{\hspace{20pt}} & \hphantom{\hspace{20pt}} & \hphantom{\hspace{20pt}}\\[-15pt]
	\ 1 & \ 3 & \ 1\\
	\ 0 & \ 0 & \ 0\\
	-1 & -3 & -1
	\end{bmatrix}
	y:
	\begin{bmatrix}
	\hphantom{\hspace{20pt}} & \hphantom{\hspace{20pt}} & \hphantom{\hspace{20pt}}\\[-15pt]
	\ 3 & \ 6 & \ 3\\
	\ 0 & \ 0 & \ 0\\
	-3 & -6 & -3
	\end{bmatrix}
	y+1:
	\begin{bmatrix}
	\hphantom{\hspace{20pt}} & \hphantom{\hspace{20pt}} & \hphantom{\hspace{20pt}}\\[-15pt]
	\ 1 & \ 3 & \ 1\\
	\ 0 & \ 0 & \ 0\\
	-1 & -3 & -1
	\end{bmatrix}
	\\
	z-1:
	\begin{bmatrix}
	\hphantom{\hspace{20pt}} & \hphantom{\hspace{20pt}} & \hphantom{\hspace{20pt}}\\[-15pt]
	-1 & \ 0 & \ 1\\
	-3 & \ 0 & \ 3\\
	-1 & \ 0 & \ 1
	\end{bmatrix}
	z:
	\begin{bmatrix}
	\hphantom{\hspace{20pt}} & \hphantom{\hspace{20pt}} & \hphantom{\hspace{20pt}}\\[-15pt]
	-3 & \ 0 & \ 3\\
	-6 & \ 0 & \ 6\\
	-3 & \ 0 & \ 3
	\end{bmatrix}
	z+1:
	\begin{bmatrix}
	\hphantom{\hspace{20pt}} & \hphantom{\hspace{20pt}} & \hphantom{\hspace{20pt}}\\[-15pt]
	-1 & \ 0 & \ 1\\
	-3 & \ 0 & \ 3\\
	-1 & \ 0 & \ 1
	\end{bmatrix}
 \end{split}
 \label{Sobel3D}
\end{equation}
\end{figure*}

Der Operator wird ähnlich wie der Central-Difference-Operator angewandt. Die Operatormaske wird jeweils in der X-, in der Y- und in der Z-Ebene entsprechend einen Voxel vor dem Abtastpunkt, genau auf dem Abtastpunkt und einen Voxel nach dem Abtastpunkt angelegt. Die Zahlen in den Operatormasken geben Koeffizienten an, mit denen der Wert des Voxels an der entsprechenden Position multipliziert werden muss. Alle so errechneten Werte einer jeden Maske werden für jede Ebene danach addiert und ergeben so eine Komponente des Gradienten. Da sowohl der Sobel3D-Operator als auch der Zucker-Hummel-Operator deutlich mehr Voxel in die Gradientenberechnung mit einbeziehen, gibt der errechnete Gradient eine deutlich bessere Beschreibung der lokalen Oberflächenstruktur im Volumendatensatz wieder als es der Central-Difference-Operator vermag.

\subsection{Bounding Box}
\begin{figure}[htbp]
\begin{center}
  \includegraphics[width=0.5\columnwidth]{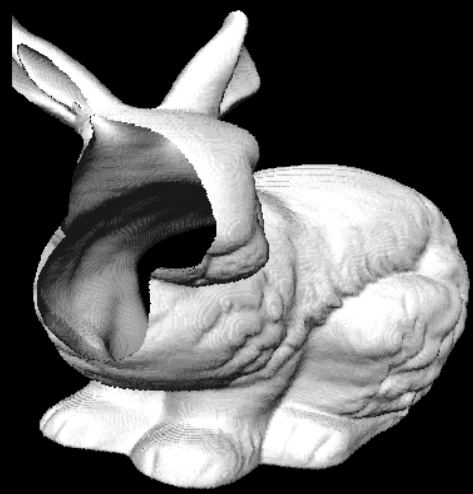}
  \caption[Bounding Box Stanford Bunny]{Bounding Box Stanford Bunny}
  \label{label:stanford_bunny_aufgeschnitten}
\end{center}
\end{figure}

Wie bereits erwähnt wird eine Bounding Box genutzt, um den Strahleintritt in den Volumendatensatz und den Strahlaustritt aus dem Volumendatensatz zu bestimmen. Dies beschleunigt zum einen die Berechnung, da Strahlen die ``am Volumen vorbeigehen'' nicht abgetastet werden müssen. Zum anderen ermöglicht die Bounding Box auch das ``Zerschneiden'' des Volumens. So kann z.B. in einen Datensatz ``hineingeschaut'' werden. In Abbildung \ref{label:stanford_bunny_aufgeschnitten} sieht man ein Computertomographie-Bild des berühmten \textit{Stanford Bunny} dessen Bounding Box so gesetzt wurde, dass man in den Volumendatensatz hinein und sogar hindurch schauen kann. Hier kann man erkennen, dass der original Porzellanhase einen Hohlraum beinhaltet. Die scheinbar nicht vorhandene Dicke der ``Hülle'' das Hasen liegt an der Berechnung der Gradienten. Die Strahlenberechnung beginnt zwar ``innerhalb'' des Hasenkörpers, jedoch werden für die Gradientenbestimmung auch Daten vor Strahlbeginn genutzt. Somit ist der Gradient bei Strahlbeginn 0 da der Strahl ``mitten im Material'' beginnt.

\subsection{Farbbestimmung}
Um den Schnittpunkten nicht nur eine Schattierung (Shading) sondern auch eine natürliche Farbe zu geben müssen den Volumendaten Farbwerte zugeordnet werden. Entweder beinhaltet das Volumen bereits Farbwerte die jedem Voxel mitgegeben werden. Oder, wie im Fall von Computertomographie, beinhalten die Volumendaten allerdings andere Werte, wie z.B. die ortsabhängige Abschwächung der Röntgenstrahlung durch Absorption im Gewebe. Allerdings kann man aus diesen Werten Farbwerte ableiten/zuweisen. So lässt sich mit Hilfe der sogenannten \textit{Hounsfield-Skala} \cite{hounsefield} einem jeden Voxelpunkt auch ein Farbwert zuweisen:

\begin{equation}
  Hounsfield(\mu_{Gewebe}) = \frac{\mu_{Gewebe} - \mu_{Wasser}}{\mu_{Wasser}} * 1000 HU
\end{equation}

Mit Hilfe dieser Formel kann aus dem Abschwächungskoeffizienten des Gewebes ($\mu_{Gewebe}$) ein Hounsfield-Wert bestimmt werden. Diesen Hounsfield-Werten wiederum lassen sich Farbwerte zuweisen. Wasser besitzt, durch die Normierung in der Formel, einen HU-Wert von 0. Luft besitzt einen Wert von -1000 HU. Gewebe besitzt einen Wert um die -100 HU und Knochen einen zwischen 500 und 1500HU. Hier kann man Gewebe rötliche Töne, Knochen weiße und Luft schwarze Werte zuweisen. Auch ist es dank dieser Skala möglich, bestimmte Gewebetypen (z.B. Krebsgewebe) hervorzuheben (einzufärben) oder ganz auszublenden. Diese Farbskalen werden häufig als Lookup Tabelle implementiert, in der zu einem HU-Wert ein Farbwert ausgelesen werden kann.

\subsection{Beleuchtung}
Damit die Oberflächen des Volumendatensatzes verschiedene Schattierungen erhalten (Shading), wird ein einfaches Beleuchtungsmodell mit nur einer einzigen Lichtquelle benutzt. Die Lichtquelle wird als einfache Punkt-förmige Quelle modelliert, die durch den Vektor $lightPosition$ beschrieben wird. Um den Beleuchtungswert im Schnittpunkt zwischen Strahl und Oberfläche zu bestimmten wird außerdem der Gradient $norm$ im Schnittpunkt benötigt. Mittels des Skalarprodukts zwischen Normalenvektor im Schnittpunkt (Gradienten) und der normalisierten Differenz aus Lichtposition und Schnittpunkt wird die Stärke der Beleuchtung bestimmt. Im Codebeispiel \ref{label:shading} ist ein einfacher Algorithmus für die Beleuchtung in OpenCL dargestellt.

\begin{small}
\begin{lstlisting}[caption=Volume Shading, label=label:shading]{Volume Shading}
float4 Shade(float4 pos,
             float4 norm,
             float4 lightPosition)
 {
	float4 color =
	(float4)(1.0f,1.0f,1.0f,0.0f);			
	//Farbe der Lichtquelle: Weiss
	
	float4 livec =
	fast_normalize(lightPosition - pos);	
	//Differenz aus Lichtposition und Schnittpunkt
	
	float illum =
	dot(livec,norm);							
	//Grad der Beleuchtung; Skalarprodukt
	
	return illum * color;									
	//Farbe zurueckgeben
 }
\end{lstlisting}
\end{small}

\begin{figure}[htbp]
\begin{center}
  \includegraphics[width=0.5\columnwidth]{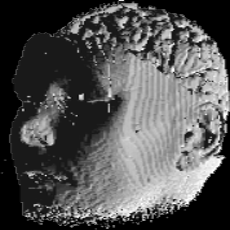}
  \caption[Beleuchtetes Gehirn 1]{Beleuchtetes Gehirn 1}
  \label{label:beleuchtung1}
  \includegraphics[width=0.5\columnwidth]{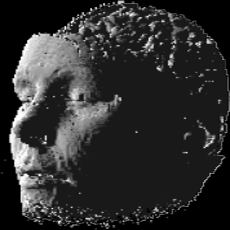}
  \caption[Beleuchtetes Gehirn 2]{Beleuchtetes Gehirn 2}
  \label{label:beleuchtung2}
\end{center}
\end{figure}

Indem die Position der Lichtquelle bewegt wird (verändern des Vektors $lightPosition$) lässt sich das zu raycastende Bild unterschiedlich ausleuchten. In Abbildungen \ref{label:beleuchtung1} und \ref{label:beleuchtung2} ist eine Abbildung eines menschlichen Kopfes mit freigelegten Gehirn mit zwei unterschiedlichen Beleuchtungspositionen ($lightPosition = (-1000,-1000,-1000)$ und $lightPosition = (1000,1000,1000)$) abgebildet.

\subsection{Transparenz-Formel}
\cite{Pawasauskas} zeigt eine Transparenz-Formel, mit der sich Voxel, die hintereinander liegen und mittels eines Strahls abgetastet werden, so verrechnet werden können, das auch in das Volumen hinein geschaut werden kann:

\begin{equation}
	C_out = C_in (1 - \alpha(X_i )) + C(X_i )\alpha(X_i)
\end{equation}

Wobei $C(x)$: eine Schattierung, errechnet aus dem lokalen Gradienten ist, $\alpha(X)$ die Lichtundurchlässigkeit, berechnet aus dem CT-Wert des Voxels ist, $C_out$ die ausgehende Intensität/Farbe des aktuellen Voxels ist und $C_in$ die eingehende Intensität in den Voxel ist. Diese Werte errechnen sich zum einen aus den Daten der Voxel bzw des Volumendatensatzes und zum anderen aus der Lichtquelle, mit der der Volumendatensatz beleuchtet werden soll. Diese Formel wird rekursiv so lange entlang des Strahls angewandt, bis auf ein Alpha-Wert innerhalb des Volumens getroffen wird, der 0 ist (z.B. durch Treffen auf lichtundurchlässiges Gewebe wie Knochen). Der in dieser Ausarbeitung entwickelte Raycaster implementiert die Transparenz-Formel jedoch nicht und bietet somit nur Bilder ohne Transparenz.

\subsection{Gesamtaufbau eines Raycasters}
Der grundlegende Raycasting-Algorithmus wird mittels der oben beschriebenen Techniken implementiert und lässt sich wie folgt darstellen:

\begin{enumerate}
 \item \textbf{Raycasting:} Ausgehend vom Auge des Betrachters wird durch jeden Bildpunkt des zu rendernden Bildes ein Strahl geleitet. Der Strahl trifft auf das Volumen, dass abgetastet werden soll und verlässt dieses wieder. Eine BoundingBox kann als einfaches Mittel dienen, um die Berechnungsdauer zu verkleinern. So wird das Volumen innerhalb der Box platziert um so den Startpunkt und Endpunkt der folgenden Abtastung zu bestimmen. So erspart man sich die Abtastung von ``leeren'' Raum.
 \item \textbf{Sampling:} Entlang des im ersten Schritt bestimmten Strahls werden Sampling-Punkte gewählt und das Volumen wird ``abgetastet''. Da Voxel nicht exakt getroffen werden dienen Techniken wie lineare oder trilineare Interpolation dazu, bessere Abtastergebnisse zu erhalten.
 \item \textbf{Shading:} Für jeden bestimmten Abtastpunkt wird mittels Gradientenoperator ein Gradient bestimmt. Dieser Gradient gibt die lokale Struktur der Oberfläche des im Volumendatensatz befindlichen Objektes wieder Mittels dieses Gradienten und einer gegebenen Lichtquelle lassen sich die Beleuchtung und Farbe der Punkte ermitteln.
 \item \textbf{Compositing:} Nachdem alle Abtastpunkte berechnet und ihre Beleuchtung bestimmt wurde lässt sich mittels eines Transmissions-Emmissions-Modells der Einfluss eines jeden Punktes auf den zu berechnenden Bildpunkt bestimmen um so eine resultierende Farbe zu bestimmen. Diese wird letzendlich an die zu berechnende Stelle im Bild gezeichnet.
\end{enumerate}

\subsection{Implementierung in .NET und OpenCL}

Der für diese Ausarbeitung implementierte Raycaster ist mit Hilfe von Microsoft .NET C\# und OpenCL unter Microsoft Windows implementiert. Ein Video, dass alle Funktionen des Raycasters demonstriert kann unter \cite{Kopal} betrachtet werden. Das Host Programm besteht zum einen aus vier Laderoutinen, die vier unterschiedliche Volumendatensätze in den Speicher der Grafikkarte laden können (Stanford Bunny, Head, Brain, Stent).\\
\\
Der Benutzer hat jederzeit die Möglichkeit, den darzustellenden Volumensatz zu ändern aber auch die Darstellung selbst zu manipulieren. Dafür gibt es auf der linken Seite der Applikation ein Settings-Fenster. Hier kann aber auch die BoundingBox abgeändert werden, um so Einblicke innerhalb des Volumens zu erhalten. Neben der BoundingBox  können auch zwei Schwellenwerte festgelegt werden, die angeben, von welchem Startwert bis zu welchem Zielwert der Hounsefield Skala Voxel überhaupt dargestellt werden sollen. So kann z.B. beim ``Head'' das Fleisch entfernt werden, um so den freigelegten Schädel darzustellen. \\
\\
\begin{figure}[htbp]
\begin{center}
  \includegraphics[width=0.8\columnwidth]{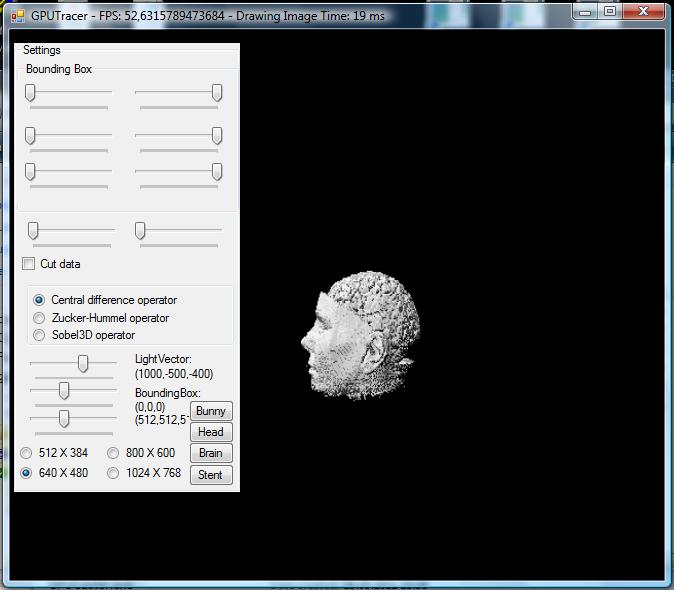}
  \caption[Implementierung eines Raycasters in .NET und OpenCL]{Implementierung eines Raycasters in .NET und OpenCL}
  \label{label:Implementierung}
\end{center}
\end{figure}
\\
Mittels weiterer Regler kann der Beleuchtungsvektor verändert werden. Er gibt die Position der einzigen Lichtquelle an, welche den Volumendatensatz ausleuchtet.\\
\\
Um auch auf langsameren Grafikkarten noch annehmbare Resultate zu erhalten, ermöglicht die Implementierung ein Umschalten der Render-Auflösung (512X384, 640X480, 800X600, 1024X768) und ein Umschalten des Gradienten Operators (Central-Difference-Operator, Zucker-Hummel-Operator, Sobel3D-Operator).\\
\\
Mit Hilfe der Maus sowie der Pfeiltasten der Tastatur lässt sich das dargestellte Model drehen sowie vergrößern und verkleinern (Zoomen). \\
\\
In der Kopfleiste der Applikation lassen sich die aktuelle Bildrate (Frames per Second) sowie die Berechnungsdauer eines einzelnen Bildes nachvollziehen.\\
\\
Der Raycaster ist, bis auf die Nutzung der BoundingBox, wenig optimiert worden. Die Abtastung innerhalb des Volumendatensatzes trickst ein wenig, indem sie zunächst in großen Schritten entlang des Strahls vorwärts abtastet, bis sie auf ``feste Materie'' trifft. Um den Eintrittspunkt genauer zu bestimmen wird ab diesem Punkt rückwärts in kleineren Schritten abgetastet, um so den Eintrittspunkt in die feste Materie zu bestimmen. Obwohl der Raycaster wenig optimiert ist, liefern selbst ältere Computer ( hier getestet mit einem Intel Core 2 Duo 6420 mit 2,1GHz und einer NVidia GForce 8800 GT ) annehmbare Bildraten im Bereich zwischen fünf und 50 Bildern pro Sekunde. Im nachfolgendem Kapitel sind einige Messungen aufgeführt, die die Performance des Raycasters darstellen und einige der implementierten Gradientenoperatoren miteinander vergleichen. Weitere Optimierungsmöglichkeiten für einen Raycaster zeigt das Kapitel ``Beschleunigungsmöglichkeiten für Raycasting''.

\section{Raycaster Evaluation}
In diesem Kapitel wird der in dieser Ausarbeitung entwickelte Raycaster bezüglich seiner Rendergeschwindigkeit untersucht. Hierzu wurde gemessen, wie viele Bilder pro Sekunde mittels des Raycasters gerendert werden können. Vergleichend sind nachfolgend unterschiedliche Auflösungen sowie unterschiedliche Gradientenoperatoren dargestellt.

\begin{table}[htbp]
\begin{center}
\begin{tabular}{|c|c|c|c|c|}
\hline & \textbf{512X384}   & \textbf{640X480}   & \textbf{800X600}   & \textbf{1024X768} \\
\hline \textbf{Bunny} & 20,00 fps & 16,13 fps & 12,65 fps & 8,92 fps \\
\hline \textbf{Head}  & 27,02 fps & 24,39 fps & 20,40 fps & 14,08 fps \\
\hline \textbf{Brain} & 30,30 fps & 27,02 fps & 23,25 fps & 16,39 fps \\
\hline \textbf{Stent} & 15,87 fps & 12,56 fps &  9,90 fps &  7,46 fps \\
\hline
\end{tabular}
\caption[Messung Central-Difference-Operator]{Messung Central-Difference-Operator}
\begin{tabular}{|c|c|c|c|c|}
\hline & \textbf{512X384}   & \textbf{640X480}   & \textbf{800X600}   & \textbf{1024X768} \\
\hline \textbf{Bunny} & 10,10 fps & 7,57  fps &  5,84 fps &  4,09 fps \\
\hline \textbf{Head}  & 18,51 fps & 15,38 fps & 11,62 fps &  7,87 fps \\
\hline \textbf{Brain} & 23,25 fps & 19,60 fps & 16,39 fps & 11,76 fps \\
\hline \textbf{Stent} &  6,53 fps &  4,90 fps &  3,62 fps &  2,69 fps \\
\hline
\end{tabular}
\caption[Messung Zucker-Hummel-Operator]{Messung Zucker-Hummel-Operator:}
\end{center}
\end{table}

\begin{figure}[htbp]
\begin{center}
  \includegraphics[width=1\columnwidth]{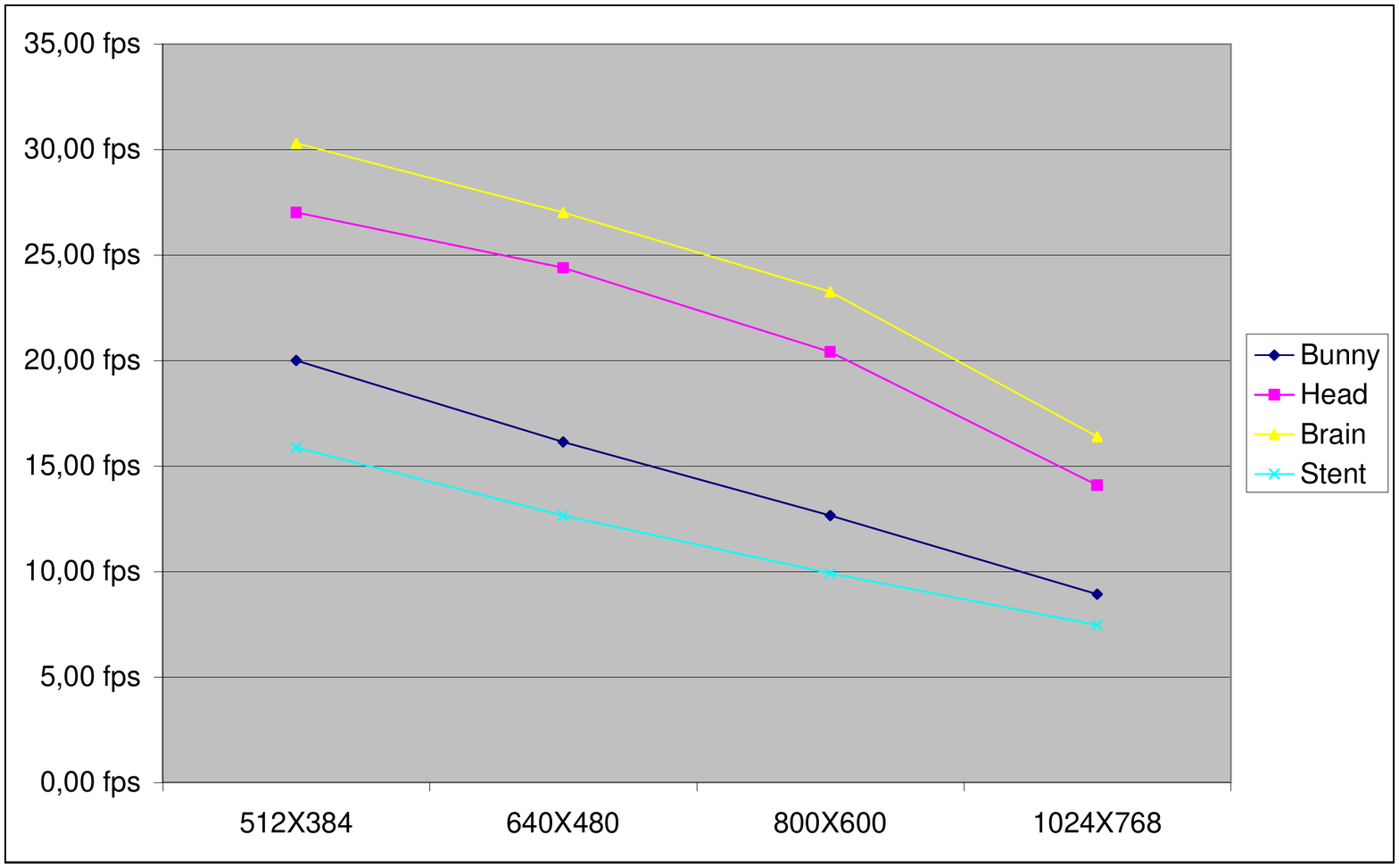}
  \caption[Central-Difference-Operator - Messungen]{Central-Difference-Operator - Messungen}
  \label{label:MessungCentralDifference}
  \includegraphics[width=1\columnwidth]{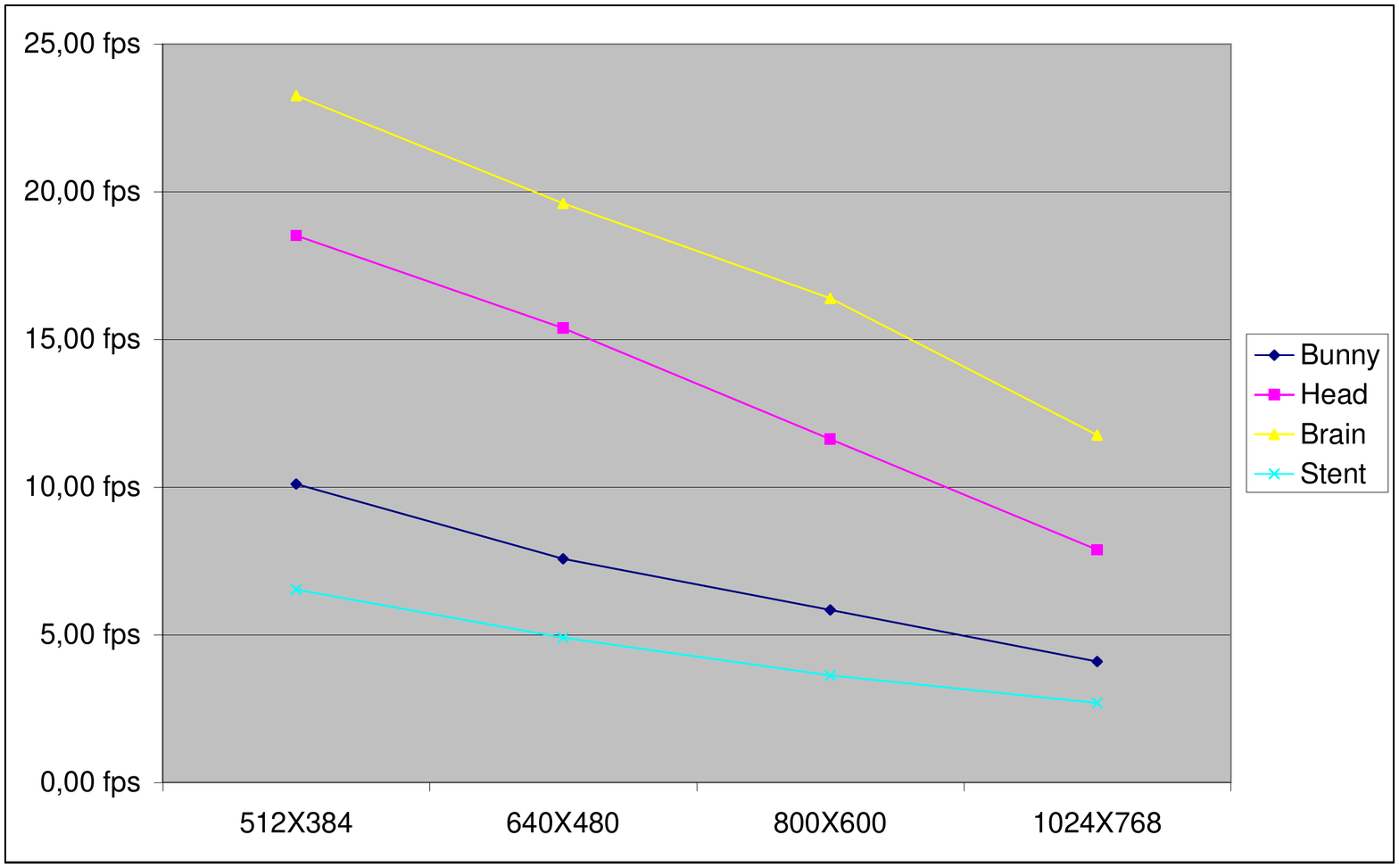}
  \caption[Zucker-Hummel-Operator - Messungen]{Zucker-Hummel-Operator - Messungen}
  \label{label:MessungZuckerHummel}
  \includegraphics[width=1\columnwidth]{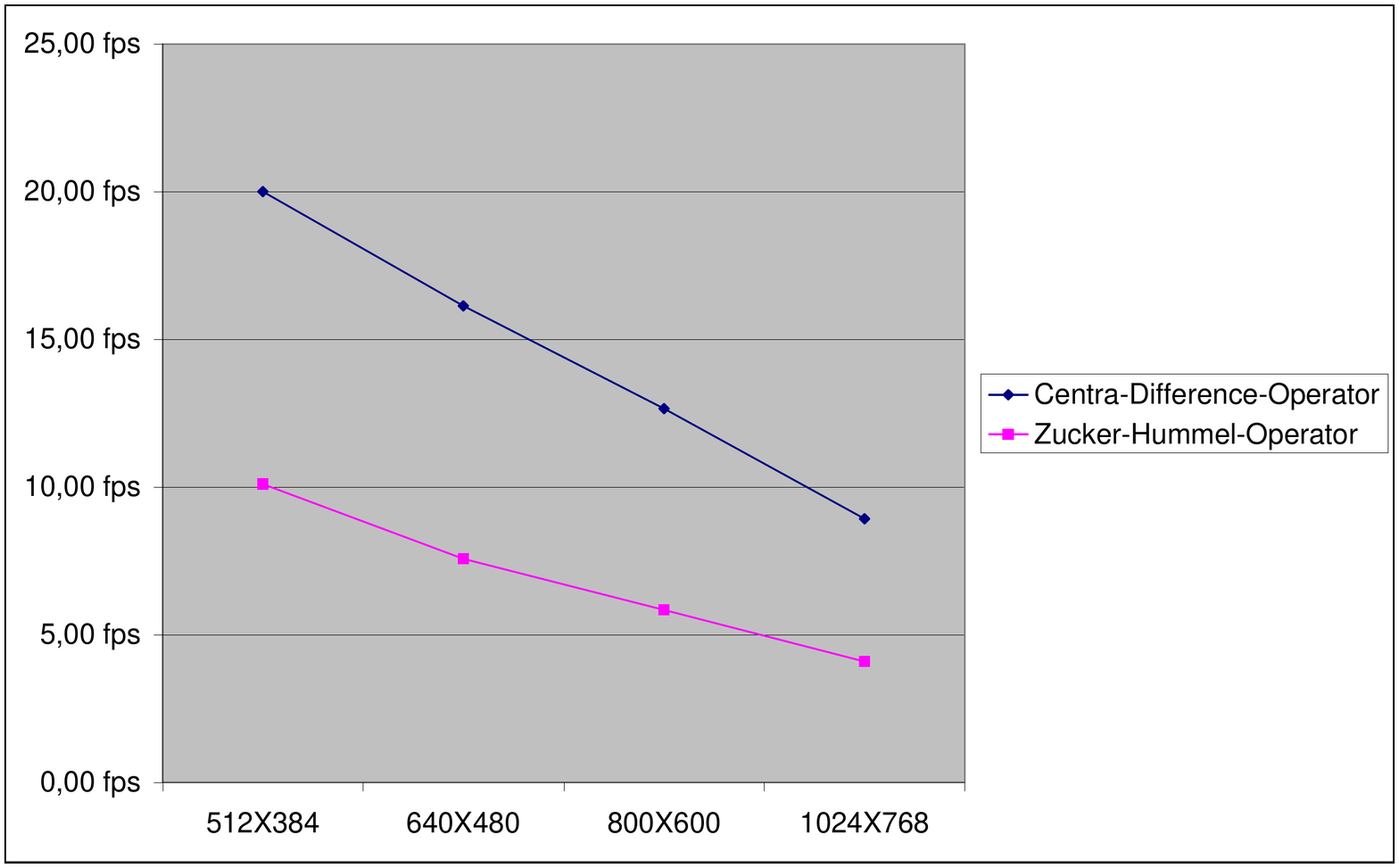}
  \caption[Direkter Vergleich Central-Difference und Zucker-Hummel - Messungen]{Direkter Vergleich Central-Difference und Zucker-Hummel - Messungen}
  \label{label:MessungVergleich}
\end{center}
\end{figure}

Die Messungen wurden mittels eines Intel Core 2 Duo 6420 mit 2,1GHz und einer NVidia GForce 8800 GT vorgenommen. Die erste Tabelle zeigt die Bildwiederholfrequenz (fps) gemessen beim Central-Difference-Operator, die zweite Tabelle beim Zucker-Hummel-Operator. Bildauflösungen sind Spaltenweise aufgetragen, die gerenderten Modelle Zeilenweise. In den Abbildungen \ref{label:MessungCentralDifference} und \ref{label:MessungZuckerHummel} sind die Daten als Graph dargestellt. Hier lässt sich ein linearer Zusammenhang zwischen Renderauflösung und Bildwiederholfrequenz erkennen. Dies ist aber auch nicht verwunderlich, da beim Erhöhen der Renderauflösung auch die Anzahl der zu berechnenden Strahlen steigt. Da pro neu zu berechnendem Pixel ein neuer Strahl entsteht und der Raycaster weder Reflexion noch Refraktion berechnet, steigt der Rechenaufwand linear. In der Abbildung \ref{label:MessungVergleich} sind beide Messungen in einer Zeichnung als Vergleich dargestellt. Der Central-Difference-Operator liefert eine deutlich höhere Bildwiederholfrequenz als der Zucker-Hummel-Operator. Dies lässt sich einfach dadurch erklären, dass der Zucker-Hummel-Operator deutlich komplexer in seiner Berechnung ist als der Central-Difference-Operator. Während der Central-Difference-Operator nur 6 Sampling-Punkte für seine Berechnung benötigt sind es beim Zucker-Hummel-Operator 54. Der Performance-unterschied des Operators wird aber wieder durch eine deutlich gesteigerte Bildqualität wett gemacht.

\section{Beschleunigungsmöglichkeiten für Raycasting}
In diesem Kapitel werden einige Techniken vorgestellt, mit denen Raycaster optimiert werden können. Mittels dieser Optimierung lassen sich die Rendergeschwindigkeit und damit die Bildwiederholfrequenz des Raycasters deutlich verbessern.

\subsection{Hitpoint-Refinement}
Beim \textit{Hitpoint-Refinement} wird ohne Einbruch der Rendering-Geschwindigkeit eine höhere Qualität im Detektieren von Oberflächen (Schnittpunkt zwischen Strahl und Oberfläche) erreicht. Zunächst wird entlang eines Strahls abgetastet. Diese Abtastung wird mit einer relativ kleinen Abtastrate durchgeführt (große Schritte zwischen den Abtastpunkten). Trifft man auf einen Schnittpunkt mit einem Objekt wird zunächst ein halber Schritt zurück gemacht. Nun wird überprüft, ob man sich näher am gewünschten Zielwert befindet oder sich von diesem entfernt hat. Befindet man sich näher am Zielwert, so geht man einen halben Schritt der letzten Schrittweite weiter zurück (also ein Viertelschritt). Ansonsten geht man einen solchen Schritt wieder nach vorne. Diese Vorgehen wiederholt sich 5 bis 6 mal und führt laut \cite{Scharsach} zu einem 64 fach besseren Schnittpunkt als der zuerst gewählte. Wird Hitpoint-Refinement eingesetzt kann die Abtastdistanz auf 400\% oder 500\% der originalen angehoben werden und so die Rendering-Geschwindigkeit deutlich erhöhen.

\subsection{Adaptives Sampling}
Beim\textit{ Adaptiven Sampling} unterteilt man den zu Grunde liegenden Volumendatensatz in unterschiedliche Regionen. Es gibt solche Regionen die Detailarm sind, wie z.B. großflächige Oberflächen. Dann gibt es andere Regionen die besonders Detailreich sind, wie z.B. Regionen mit sehr vielen kleinen Objekten. In den Detailarmen Regionen wird die Abtastrate des Volumendatensatz herunter skaliert, da eine gröbere Abtastung in diesen ausreichend ist. In den Detailreichen Regionen wiederum wird die Abtastrate entsprechend erhöht, um auch filigrane Details in die Berechnung einfließen zu lassen. Da Volumendatensätze aber häufig zumeist aus Detailarmen Regionen (im einfachsten Fall leerer Raum) bestehen, führt das adaptive Sampling zu einer deutlichen Steigerung der Rendergeschwindigkeit.

\subsection{Empty-Space-Skipping}
Das \textit{Empty-Space-Skipping} versucht, wie der Name bereits vermuten lässt, Empty-Space, also freien Raum, zu überbrücken. Die größte Rechenzeitverschwendung bei einem Raycaster ist das Abtasten ``im Leeren''. Der Raycaster verfolgt einen Strahl und tastet diesen sukzessiv ab. Abtastpunkte werden in immer gleichen Abständen gesetzt und der abgetastete Wert wird für den resultierenden Pixel ein berechnet. Durchläuft der Strahl größtenteils leeren Raum, also Positionen im Volumendatensatz die keine oder keine relevanten Daten enthalten, so wird dieser Raum dennoch abgetastet. Diese Abtastung kostet natürlich Rechenzeit. Empty-Space-Skipping versucht genau die Abtastung dieses leeren Raumes zu minimieren. So wird das im Volumen enthaltene Objekt möglichst genau durch eine ``Hülle'' aus primitiven (Dreiecken, Quads oder dergleichen) umschlossen. Nun muss nicht mehr der komplette ``Würfel'' des Volumendatensatzes abgetastet werden sondern nur das Stück des Strahls, das vom Eintrittspunkt des Strahls und der Hülle bis zu dessen Austrittspunkt reicht.

\subsection{Octree}
\begin{figure}[htbp]
\begin{center}
  \includegraphics[width=1\columnwidth]{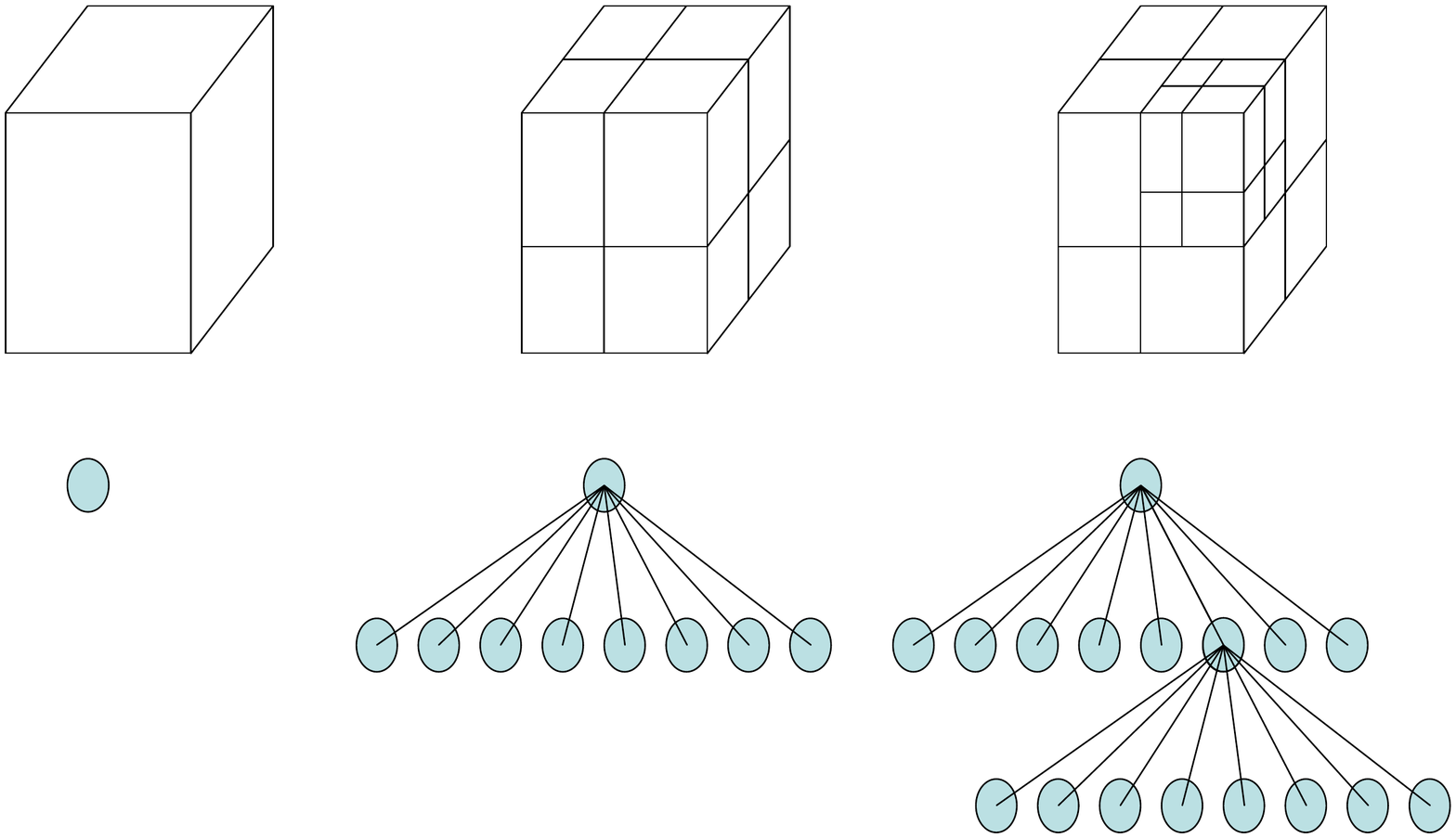}
  \caption[Aufbau eines Octrees]{Aufbau eines Octrees}
  \label{label:Octree}
\end{center}
\end{figure}

Ähnlich wie bei der Einteilung im Zweidimensionalen durch \textit{Quadtrees}, bei denen ein 2D-Bild immer wieder in vier gleich große Stücke zerteilt wird, kann ein Objekt im dreidimensionalen Raum durch einen sogenannten \textit{Octree} beschrieben werden. Hierzu wird ein Volumendatensatz in acht gleich große Blöcke zerteilt. Jeder Block, der relevante Daten enthält, wird in acht kleinere Blöcke zerteilt. Diese Blöcke wiederum werden mit ihrem Vaterknoten (dem sie umgebenden Block) verknüpft. Dieses Vorgehen wird rekursiv so lange wiederholt, bis die Blöcke eine bestimmte Mindestgröße oder die Größe eines einzelnen Voxels erreicht haben. Um für das Raycasting zu testen, welche Voxel nun von einem Abtaststrahl getroffen werden, testet man den Strahl mit dem Octree. Man fängt mit dem größten Block an und testet, ob der Strahl diesen schneidet. Wenn dies der Fall ist, testet man dessen Kinderblöcke und so weiter. Alle so gefunden Voxel werden sortiert in ein Ausgabearray geschrieben. Der Suchalgorithmus terminiert, wenn man entweder auf einen Block gestoßen ist, der keine Daten enthält oder bei den kleinsten Blöcken angekommen ist. In Abbildung \ref{label:Octree} ist der Zusammenhang zwischen Baum und Blöcken grafisch dargestellt. Hier wird ein Block immer weiter zerteilt und parallel dazu die Baumstruktur angelegt. Octrees ermöglichen das deutlich schnellere finden von Schnittpunkten in einem Volumendatensatz und tragen zu einer deutlich höhreren Abastgeschwindkeit bei, da sie, ähnlich wie das Empty-Space-Skipping dazu führen, das nur ``nicht-leerer Raum'' abgetastet wird. Auf Grafikkarten sind Octrees jedoch relativ schwierig zu implementieren, da die Programmierung auf der Grafikkarte keine rekursiven Algorithmen erlaubt und der dafür benötigtet Stack selbst implementiert werden muss. Aus diesem Grund wurde in dem hier vorgestellten Raycaster bisher kein Octree implementiert.

\section{Zusammenfassung und Ausblick}

\subsection{Zusammenfassung}
In dieser Ausarbeitung wurde das Volume Raycasting mit OpenCL dargestellt und erläutert. Es wurden zunächst die Grundlagen von Raytracing und Raycasting vorgestellt. Beim Raytracing werden Strahlen, die Rays, genutzt, um fotorealistische Bilder zu erzeugen. Hierzu werden Strahlen vom Auge des Betrachters in Richtung aller in der Szene befindlichen Objekte ausgesandt und mittels der Schnittpunkte Beleuchtung, Reflektion und Refraktion berechnet. Beim Raycasting werden ebenfalls Strahlen ausgesandt, jedoch werden diese auf sogenannte Volumendatensätze gerichtet. Reflektion und Refraktion werden beim Raycasting nicht betrachtet. Gradientenoperatoren dienen zur lokalen Oberflächenbestimmung innerhalb des Volumendatensatzes. Transmissions-Emmissionsmodelle werden genutzt, um z.B. in medizinischen Daten auch ``unter die Haut'' schauen zu können. Für die Farbbestimmung in medizinischen Daten wird die Hounsefield-Skala benutzt.

Der offene Standard OpenCL wurde vorgestellt und für diese Ausarbeitung genutzt, um einen Raycaster zu implementieren. OpenCL-Programme werden, im Gegensatz zu klassischen Computerprogrammen, direkt auf der Grafikkarte ausgeführt. Der Vorteil dieser Programme ist die massive Parallelität, die OpenCL bzw. Grafikkartenprogrammierung generelle, mit sich bringt. Aus diesem Grund bietet sich die direkte Grafikkartenprogrammierung auch an, um Raytracing- und Raycasting-Algorithmen auf diese zu portieren. Die Strahlenberechnungen sind voneinander unabhängig und können daher parallel ausgeführt werden. Hierbei ermöglichen Grafikkarten Rechengeschwindkeiten, die es dem Entwickler ermöglichen, Raytracing und Raycasting in Echtzeit ablaufen zu lassen.

Der innerhalb dieser Ausarbeitung entwickelte Raycaster und seine Funktionen wurden ebenfalls kurz vorgestellt. Anschließend zeigte eine Evalutation die Geschwindigkeiten (anhand Bildwiederholraten), die dieser relativ einfach konstruierte Raycaster, auch ohne besondere Optimierungen, liefern kann. Hierzu wurden im Evaluationskapitel zwei der implementierten Gradientenoperatoren gegenübergestellt.

Im Kapitel ``Beschleunigungsmöglichkeiten für Raycasting'' wurden mehrere Optimierungen vorgestellt, um das Raycasting ``noch schneller'' zu implementieren. Hierunter fielen Hitpoint-Refinement, Adaptives Sampling, Empty-Space-Skipping und der Octree. Allen diesen Verfahren ist gemein, dass sie unnötige Abtastungen im Volumendatensatz vermeiden bzw vermeidbar machen.

\subsection{Ausblick}
Parallel zu dieser Ausarbeitung wurde ein eigener rudimentärer Raycaster geschrieben, der allerdings relativ akzeptable Ergebnisse liefert. Nichtsdestotrotz gibt es eine noch eine Menge Optimierungen, die den Raycaster ``besser'' machen könnten. Einige dieser Optimierung wurden im Kapitel ``Beschleunigungsmöglichkeiten für Raycasting'' kurz vorgestellt. So könnte die Implementierung eines Octrees den selbst entwickelten Raycaster um einiges schneller machen. Leider machen die mangelnden Debug-Möglichkeiten und ein teilweise sehr merkwürdiges Verhalten der selbstentwickelten OpenCL Programme (z.B. Speicherüberlauf oder einfach das Abstürzen des Grafikkartentreibers) eine Entwicklung von komplexeren Algorithmen auf der Grafikkarte deutlich schwieriger. Auch unterstützt die OpenCL-Implementierung von NVidia nicht alle Funktionen, die der offene Standard mit sich bringt. So wäre eine Portierung auf das von NVidia entwickelte und vermarktete CUDA eine Möglichkeit, einen Raycaster einfacher zu entwickeln und zu Debuggen, was vor der Auswahl des dem Raycaster zu Grunde liegenden Frameworks leider nicht bekannt war. So würde eine Portierung auf CUDA vermutlich einige der angesprochenen Probleme minimieren oder beseitigen, da NVidia CUDA deutlich mehr forciert als OpenCL. Durch die Portierung auf CUDA würde der Raycaster leider seine Kompatibilität zu ATI-Grafikkarten verlieren, da es nur auf NVidia Grafikkarten arbeitet.

\bibliographystyle{IEEEtran}
\bibliography{IEEEabrv,references}

\end{document}